\documentstyle[12pt,aaspp4,psfig]{article}
\def\ltsima{$\; \buildrel < \over \sim \;$}
\def\simlt{\lower.5ex\hbox{\ltsima}}
\def\gtsima{$\; \buildrel > \over \sim \;$}
\def\simgt{\lower.5ex\hbox{\gtsima}}
\def\kms{{\rm\,km\,s^{-1}}}

\def\msun{{\rm\,M_\odot}}

\def\s{\ifmmode \widetilde \else \~\fi}
\def\={\overline}

\def\spose#1{\hbox to 0pt{#1\hss}}

\def\etal{{\it et al.\ }}

\def\lta{\mathrel{\spose{\lower 3pt\hbox{$\mathchar"218$}}
     \raise 2.0pt\hbox{$\mathchar"13C$}}}
\def\gta{\mathrel{\spose{\lower 3pt\hbox{$\mathchar"218$}}
     \raise 2.0pt\hbox{$\mathchar"13E$}}}
\def\Dt{\spose{\raise 1.5ex\hbox{\hskip3pt$\mathchar"201$}}}    
\def\dt{\spose{\raise 1.0ex\hbox{\hskip2pt$\mathchar"201$}}}    

\def\=={\equiv}

\def\dotsfill{\leaders\hbox to 1em{\hss.\hss}\hfill}

\def\jy{{\rm\,Jy}}
\def\mJy{{\rm\,mJy}}
\def\GHz{{\rm\,GHz}}

\def\mum{{\rm \mu m}}

\newcommand{\ffffff}[1]{\mbox{$#1$}}
\newcommand{\scnd}{\mbox{\ffffff{''}\hskip-0.3em .}}

\newcommand{\minu}{\mbox{\ffffff{'}\hskip-0.3em .}}
 
\newcommand{\name}{APM 08279+5255}
\newcommand{\ly}{damped Ly$_\alpha$}
\newcommand{\mg}{\ion{Mg}{2}}

\slugcomment{To appear in the Astrophysical Journal}

\lefthead{Irwin \etal}
\righthead{\name : an ultraluminous BAL quasar}

\begin{document}

\title{\name: an ultraluminous BAL quasar at a redshift z=3.87}

\author{
Michael J. Irwin\altaffilmark{1},
Rodrigo A. Ibata\altaffilmark{2}, 
Geraint F. Lewis\altaffilmark{3,4} \&
Edward J. Totten\altaffilmark{5}}

\altaffiltext{1}
{Royal Greenwich Observatory 
Madingley Rd, Cambridge CB3 0EZ, UK \nl
Electronic mail: mike@ast.cam.ac.uk}

\altaffiltext{2}
{European Southern Observatory 
Karl Schwarzschild Stra\ss e 2, D-85748 Garching bei M\"unchen, Germany \nl
Electronic mail: ribata@eso.org}

\altaffiltext{3}
{Astronomy Department, University of Washington Box 351580, 
Seattle WA 98195-1580, U.S.A. \nl
Electronic mail: gfl@astro.washington.edu}

\altaffiltext{4}
{Department of Physics \& Astronomy, University of Victoria 
PO Box 3055, Victoria, B.C. Canada V8W 3P7 \nl
Electronic mail: gfl@uvastro.phys.uvic.ca }

\altaffiltext{5}
{Department of Pure \& Applied Physics, The Queen's University of Belfast
Belfast BT71NN, Northern Ireland \nl
Electronic mail: e.totten@queens-belfast.ac.uk }

\begin{abstract}
We report on the discovery of a highly luminous, broad absorption line
quasar at a redshift of $z=3.87$ which is positionally coincident,
within one arcsecond, with the IRAS FSC source F08279+5255. A chance
alignment of the quasar and the IRAS source is extremely unlikely and
we argue that the optical and FIR flux are different manifestations of
the same object. With an R-band magnitude of 15.2, and an IRAS
60$\mum$ flux of $0.51\jy$, \name\ is (apparently) easily the most
intrinsically luminous object known, with ${\rm
L_{Bol}\sim5\times10^{15}L_{\odot}}$.  Optical CCD photometry of the
system, taken in good seeing, shows evidence that the system is
slightly elongated.  Though this data is consistent with the
superposition of the quasar on a vastly luminous galaxy, we argue that
a more likely scenario is that the optical image implies the presence
of two unresolved point-sources.  Such a configuration suggests that
gravitational lensing may play a significant role in amplifying the
intrinsic properties of the system. Point-spread-function fitting of
two discrete sources gives a separation of $\sim 0\scnd4$ and an
intensity ratio $\sim 1.1$.  The optical spectrum of the quasar
clearly reveals the presence of three potential lensing galaxies, \mg\
absorption systems at $z=1.18$ and $z=1.81$, and a \ly\ absorption
system at $z=3.07$. Additional, as yet unseen, lensing galaxies may
also be present.  We estimate the total amplification of the optical
component to be $\approx40$, but, due to the larger scale of the
emitting region, would expect the infrared amplification to be
significantly less.  Even making the conservative assumption that all
wavelengths are amplified by a factor 40, \name\ still possesses a
phenomenal luminosity of ${\rm \simgt 10^{14}L_{\odot}}$, indicating
that it belongs to a small, but significant population of
high--redshift, hyperluminous objects with copious infrared emission.
\end{abstract}


\newpage

\section{Introduction}\label{introduction}
Apart from the detection of two strongly gravitationally lensed
objects, the Seyfert II-like object IRAS F10214+4724 with ${\rm z =
2.286}$, and the broad absorption line (BAL) quasar H1413+117 -- ``the
cloverleaf'' -- at ${\rm z = 2.558}$, and the recently identified
ultraluminous galaxy SMM 02399-0136 at ${\rm z = 2.8}$, which is
mildly lensed by a galaxy cluster, there have been no detections of
far infrared (FIR) emission over the wavelength range
10$\mum$--100$\mum$ in objects with a redshift ${\rm z > 2}$.
F10214+4724 was discovered as part of a systematic redshift survey
covering an area of 700 square degrees, of 1400 IRAS {\it Faint Source
Catalogue} (FSC) galaxies brighter than $0.2\jy$ at $60\mum$ by
\cite{ro91}. SMM~02399-0136 was similarly identified in a
submillimeter survey of the fields of galaxy clusters exhibiting
strong gravitational lensing features (\cite{iv98}), appearing as the
brightest source at 450$\mum$ in the field of Abell 370.  H1413+117,
which had previously been identified as a gravitationally lensed BAL
quasar (\cite{ma88}), was found to be an IRAS FSC entry by
\cite{ba95}, after it had been detected in the submillimeter continuum
and had been shown to have CO line emission (\cite{ba92,ba94})
suggestive of thermal radiation from dust.  Further recent detection
of CO emission by \cite{om96} and \cite{gu97} in two even higher
redshift quasars, ${\rm z > 4}$, testify to the likely ubiquitous
presence of dust at early stages of galaxy evolution.  Neither of
these ${\rm z > 4}$ quasars has a counterpart in the IRAS FSC.  Indeed
it is clear that without the amplification due to gravitational
lensing ($\approx \times 30$) both F10214+4724 and H1413+117 would
also be absent from the IRAS FSC.

Due to the large number of sources in the combined IRAS {\it Point
Source Catalog} (PSC) and FSC databases, spectroscopic followup
campaigns have of necessity been rather limited in scope.
Complementary to the FSC redshift survey of Rowan-Robinson \etal 
and the more recent extension of this by \cite{ol96}, were
the ``all-sky'' followup surveys of IRAS sources based on the PSC.
The vast majority of these identifications belonged to low redshift,
${\rm z < 0.1}$, galaxies.  Notable among these were the surveys of
\cite{ro90} who carried out a sparse-sampled redshift survey of 2163
IRAS PSC $60\mum$ sources with $F_{60} \ge 0.6\jy$ at $|b| >
10^\circ$; and the IRAS galaxy survey of \cite{st90} also flux limited
at 60$\mum$ with $F_{60} \ge 1.2\jy$ at $|b| > 5^\circ$.  Apart from
the Rowan-Robinson \etal and Oliver \etal FSC surveys there appear to 
have been no systematic spectroscopic followup campaigns attempted for 
the FSC catalogue, which helps to explain why the significance of the FSC
source F08279+5255, introduced below, had not been evident before.

While conducting a survey for distant cool carbon stars in the
Galactic Halo we discovered an extremely bright 15${\rm ^{th}}$
magnitude BAL quasar at a redshift ${\rm z = 3.87}$.  The quasar is
positionally coincident, within an arcsecond, with the IRAS FSC object
F08279+5255.  Further ground-based imaging and spectroscopy of other
potential optical counterparts within 1 arcmin of the IRAS source
revealed no other likely optical counterpart.  The excellent
positional coincidence, extreme optical rarity of such bright high
redshift quasars and the IRAS spectral signature, provide a compelling
case that the IRAS source is related to the extreme properties of the
BAL quasar.

In the remainder of this Letter we briefly discuss the survey leading
to the discovery of the BAL quasar; the current suite of observations
we have made to determine its nature; discuss various hypotheses
regarding the object; and speculate on the likelihood of further
surprises in the IRAS object catalogue databases.

\section{Observations}\label{observations}

\subsection{Spectroscopic observations}\label{spectra}

The spectroscopic observations were taken as part of the APM survey to 
identify distant cool carbon stars in the Galactic Halo 
using the 2.5m Isaac Newton 
Telescope (INT) at the Roque de los Muchachos Observatory, La Palma, 
between February 27 and March 6, 1998.  Given the expected colors and 
magnitudes of cool carbon stars, candidate objects with ${\rm 11<R<17}$ and 
${\rm B_J-R \simgt2.5}$ had been selected for observation from APM sky 
catalogs at Galactic latitudes $|b| > 30^\circ$ (see \cite{ed98} for more 
details).  Candidates were observed over the wavelength range 3850\AA\  -- 
7250\AA\  using the intermediate dispersion spectrograph (IDS) with the 
R300V grating and a 1k$\times$1k TEK CCD, giving a resolution of 3.3\AA/pixel.
Pseudo-real time extraction was employed to ascertain the nature of the 
candidate objects.

\begin{figure}
\centerline{
\psfig{figure=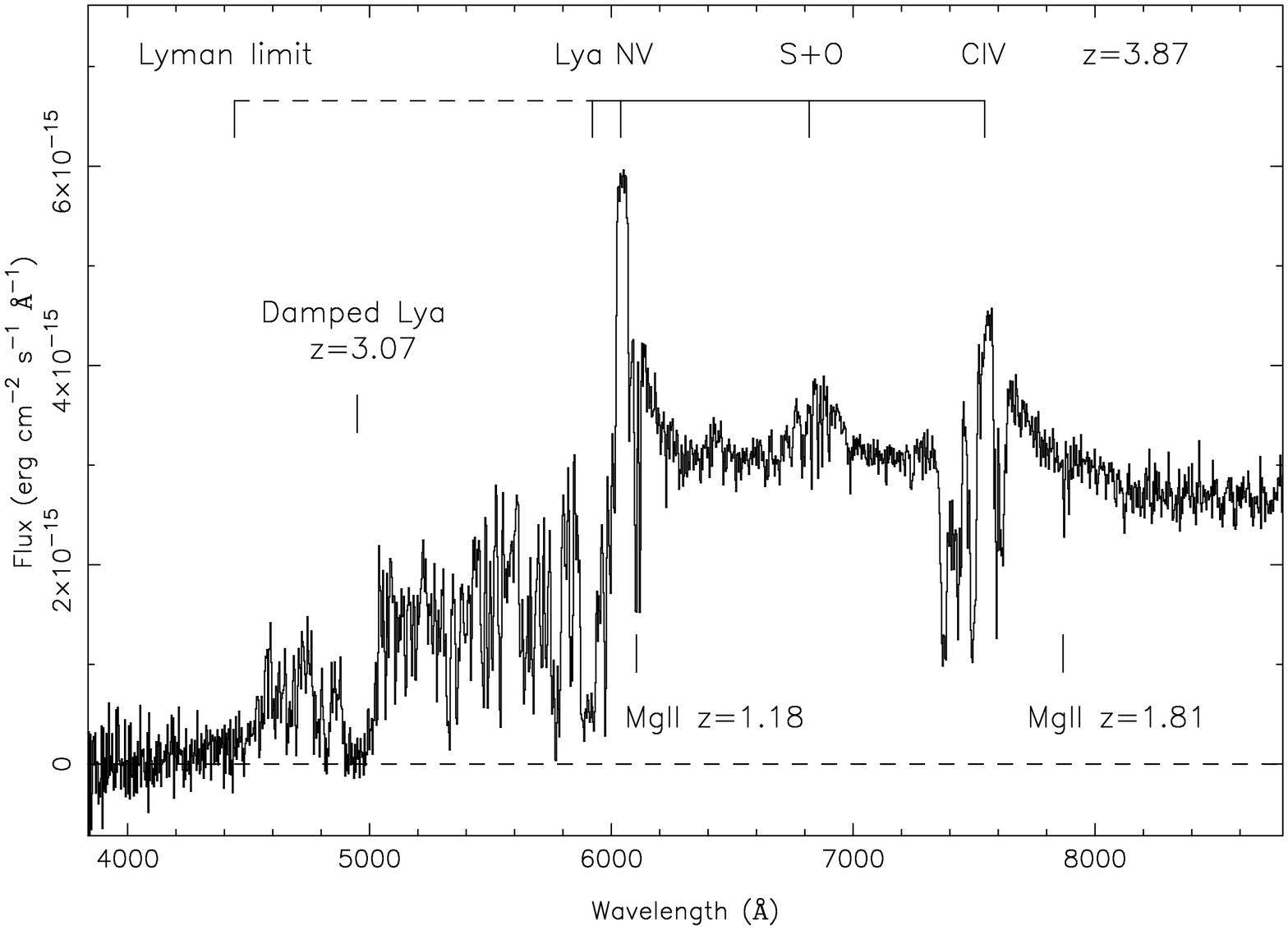,width=7in,angle=270}
}
\caption[]{  
Optical spectra of \name\ taken with the 2.5m Isaac Newton Telescope.
Features typical of high redshift quasars, namely strong \ion{N}{5} 
and \ion{C}{4}
emission lines, are clearly visible. The emission lines are seen to
possess broad absorption troughs consistent with high--velocity bulk
outflows associated with the quasar. It is interesting to note that
${\rm Ly_{\alpha}}$ emission is extremely weak, probably due to 
self-absorption by neutral gas along the line-of-sight but within the quasar.  
Also apparent is absorption beyond the Lyman limit and in the 
${\rm Ly_{\alpha}}$ forest. Coupled with this, a \ly\ system is visible 
at ${\rm z=3.07}$, as well as \ion{Mg}{2} doublets of equivalent widths
10.3\AA\ and 2.8\AA, consistent with galactic systems at ${\rm z=1.18}$
and ${\rm z=1.81}$ respectively. }
\label{fig1}
\end{figure}

One of the observed sources, at $\alpha=8^h 27^m 57.89^s$,
$\delta=52^\circ 55' 26.4''$ (B1950) with an R-band magnitude of R$=$15.2,
was clearly a highly luminous quasar at a redshift close to z$=$4.  
The spectrum displayed in Figure~\ref{fig1} is a combination of two spectra 
with overlapping wavelength coverage and shows the typical mix of broad 
emission lines and strong blueshifted absorption lines found in high 
redshift BAL quasars (\cite{tu88,we95,sl96}).  Interpretation of 
the spectrum is complicated by strong self-absorption of the Ly$_\alpha$
line and the complex pattern of absorbers blueward of \ion{C}{4}~$\lambda1549$
and close to the Ly$_\alpha +$ \ion{N}{5} lines.
However, the presence of a relatively clean S $+$ O feature 
provides a straightforward estimate of the systemic redshift, which we 
confirmed by obtaining a spectrum up to 10000\AA\ ,where 
\ion{C}{3}]~$\lambda1909$ was plainly visible at $\sim9300$\AA.  

Blueward of Ly$_\alpha$ there is a strong continuum drop caused by
intervening Ly$_\alpha$ forest absorbers with a complex series of
absorption features dominating the spectrum. A candidate damped
Ly$_\alpha$ system at z$=$3.07 is clearly present, which we confirmed
with further higher resolution spectra.  The ``grey'' Lyman-limit at
$\approx$4400\AA\ is at the redshift of the quasar and is associated
with the strong self-absorption of Ly$_{\alpha}$ seen at 5920\AA.  The
broad emission features at 6040\AA\ and 7550\AA\ can be identified
with \ion{N}{5}~$\lambda1240$ and \ion{C}{4}~$\lambda1549$, as
expected from a quasar at ${\rm z=3.87}$. Emission in ${\rm
Ly_{\alpha}}$ is apparent, but is much weaker than emission from other
species; this is a point we will return to in 
Section~\ref{conclusions_section}.

Two \ion{Mg}{2} absorbers are clearly present in the quasar spectrum.
The stronger of the two is slightly redward of \ion{N}{5}, with an
equivalent width of 10.3\AA\ and at a redshift of ${\rm z=1.18}$.
Further to the red beyond \ion{C}{4} is another \ion{Mg}{2} doublet 
with equivalent width 2.8\AA\ highlighting the presence of another 
intervening galaxy at ${\rm z=1.81}$.
We have taken higher resolution spectra over the wavelength range 4000\AA\ 
-- 10000\AA\  to search for more examples of intervening absorption systems,
but apart from the complex series of absorption lines around the main
quasar emission features we have found no evidence for further intervening
absorbers.  Although we attempted to correct for atmospheric absorption 
using a feartureless flux standard as a template, residual atmospheric A-band 
absorption features are still apparent redward of \ion{C}{4}.

\subsection{Photometry}\label{photo}

The APM measurements of the Palomar sky survey plates E/O-679 taken in
February 1953 were calibrated using the Guide Star photoelectric
sequence for this field (\cite{lask86}) and indicate that the source 
had an R-band magnitude in 1953 of ${\rm R=15.6\pm0.1}$ and a colour 
equivalent to ${\rm B_J-R=2.8\pm0.1}$.

\begin{figure}
\centerline{
\psfig{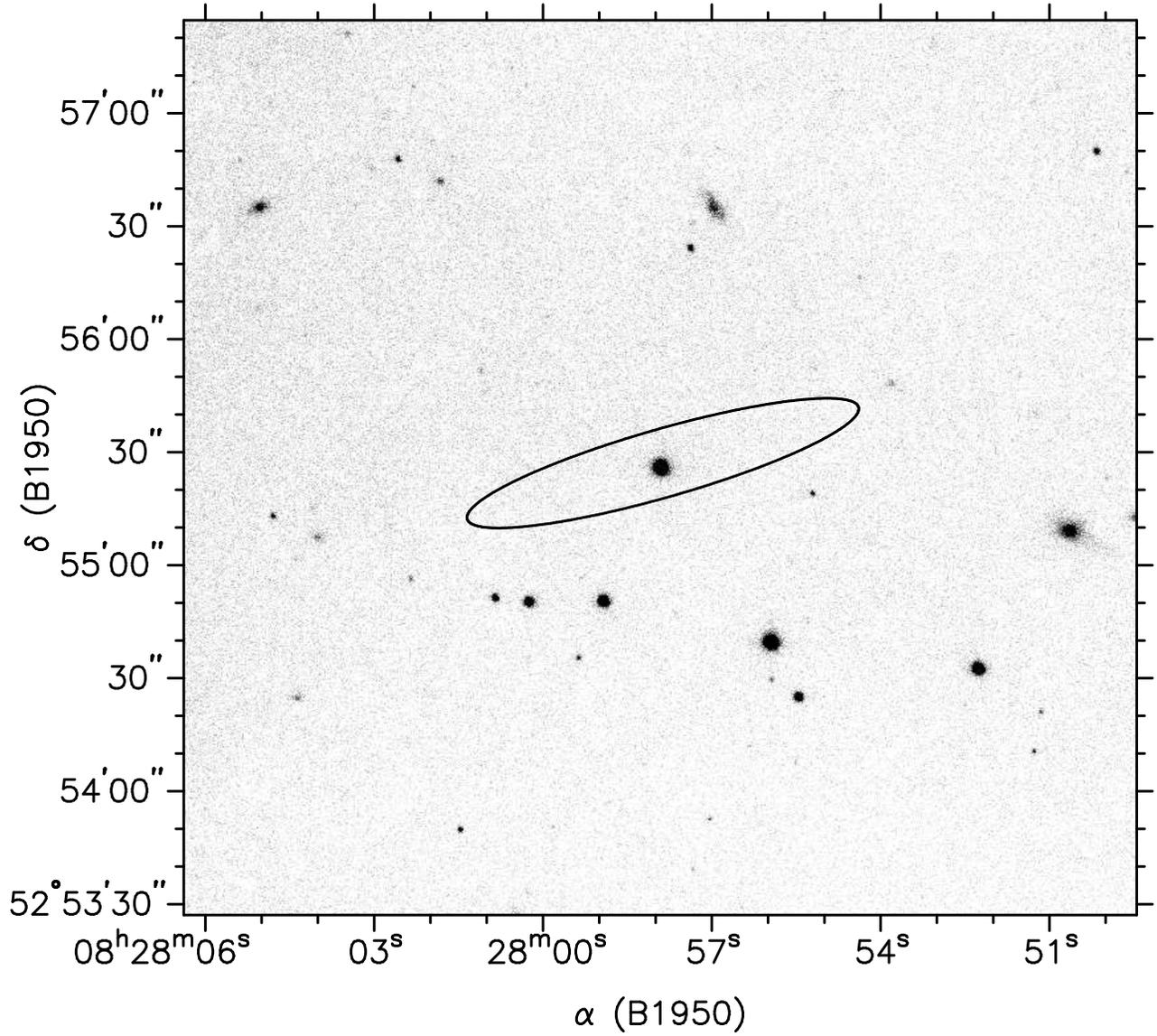}
}
\caption[]{ The $3\sigma$ uncertainty ellipse in the position of the IRAS
source F08279+5255 is shown superimposed on a $4\times 4$ arcminute
subsection of the R-band image obtained from the JKT photometry. Above the
detection limit of $R=21.4$, the only object present within this ellipse is
the ${\rm R=15.2}$ BAL quasar discovered in the APM survey.}
\label{fig2}
\end{figure}

\begin{figure}
\centerline{
\psfig{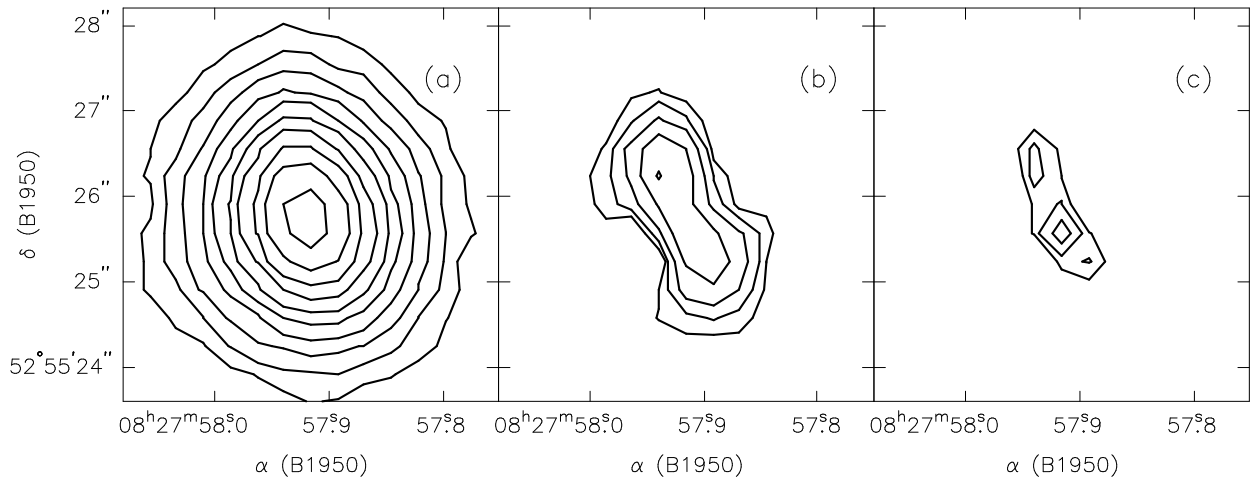}
}
\caption[]{The contours of the BAL quasar are displayed in panel (a). Panel
(b) shows the residuals resulting from fitting a single PSF to the image of
panel (a); the fitting process was constrained to give positive
residuals. The residual is rather bright, being only 2.7~magnitudes fainter
than the original image. Panel (c) shows the residuals from the best fit
with two PSFs (image separation $d=0\scnd33$, brightness ratio
$I_1/I_2=1.06$). The contour levels in all these diagrams start at ${\rm 1
e^-/pixel/s}$, and increase in equally spaced logarithmic intervals to
${\rm 100 e^-/pixel/s}$.}
\label{fig3}
\end{figure}

Following the spectroscopic detection, further photometric data were
kindly taken on our behalf by David Schlegel using the CCD camera on
the 0.9m Jacobus Kapteyn Telescope (JKT), also at the La Palma observatory.
Two frames, exposed for 150~s and 450~s in a broad-band R filter,
were obtained.  Covering $5\minu6$ on a side, these frames probe the
field around the BAL quasar with a resolution of ${\rm
0\scnd33/pixel}$.  Although the conditions were not photometric, the
seeing was measured to be $0\scnd9$.  A reproduction of the
central 4$\times$4 arcminutes of the long exposure frame is given in
Figure~\ref{fig2}, with the IRAS 3-$\sigma$ error ellipse superimosed.
Contours of the quasar image, derived from
this frame, are displayed in panel (a) of Figure~\ref{fig3}.

The raw CCD frames were pre-processed using standard techniques. The
DAOPHOT package (\cite{st87}) was employed to find and perform
aperture photometry on objects in the frame.  Relative to the field
stars, the CCD image of the quasar was found to be $0.38 \pm
0.05$~magnitudes brighter than the Palomar plate image, exposed in
1953.  Variability of this amplitude is commonly observed in quasars 
(\cite{ho94}).  Thus, a better estimate of the present brightness of the 
object is ${\rm R=15.2\pm0.1}$.

A point-spread function (PSF) for the frame was also constructed with
the DAOPHOT package; modeled as a sum of a Gaussian and a Lorentz
function plus a look-up table, this PSF was also constrained to vary
linearly over the frame.  The resulting PSF model gives an excellent
fit to $\sim 50$ point-sources over the frame, but the fit to the BAL
quasar image is poor, as the DAOPHOT goodness-of-fit parameter
$\chi=1.8$.  Imposing the constraint of non-negative residuals to the
image fitting, results in the elongated image of image residuals
displayed in Figure~\ref{fig3}b.  The brightness of this residual is
$\approx$2.7~magnitudes fainter than that of the original image.

Since the image of the quasar is elongated, it is natural to also
attempt to fit it as the sum of two slightly-offset point sources.  A
good fit results if the separation between the images is in the range
$0\scnd30 \simlt d \simlt 0\scnd45$, and if the ratio of the relative
brightnesses is in the range $1.05 \simlt I_1/I_2 \simlt 1.15$. The
residuals from the best fit solution with $\chi=1.2$ 
are displayed in Figure~\ref{fig3}c.

Although no other obvious optical counterparts are close to the IRAS 
3-$\sigma$ error ellipse, we obtained spectra of the two galaxies visible 
to the North and to the West of the quasar respectively.  Neither had
unusual features that could be attributed to the IRAS emission.

\section{Spectral Energy Distribution}\label{sed}

A search through various on-line sky catalogues showed that the BAL
quasar had been detected in both the IRAS FSC and FIRST surveys; these
are detailed in Table~\ref{table}.  The IRAS source F08279+5255, is at
$\alpha = 8^h 27^m 57.87^s$, $\delta = 52^\circ 55' 27.1''$, within an
arcsecond of the BAL, as illustrated in Figure~\ref{fig2}. There is
also a positionally coincident source in the FIRST survey at $\alpha =
8^h 27^m 57.96^s$, $\delta = 52^\circ 55' 27.2''$ with a flux of ${\rm
0.92mJy}$ at $1.4\GHz$.  However, the ROSAT Bright Source Catalog
(with a limiting countrate of ${\rm 0.05 cts/s}$ in the 0.1--2.4~keV
energy band) contains no sources within 30 armin of the
quasar. Neither is any emission detected in the NVSS catalog (at
$1.4\GHz$ with a detection limit of $2.5\mJy$) or in the WENSS catalog
(at $4.85\GHz$ with a detection limit of $18\mJy$). The lack of
significant radio flux in this system is consistent with its BAL
nature, as these systems are typically radio--quiet.  Indeed \name\
lies along the extension of the radio-quiet locus found in the
well-studied correlation between radio and FIR luminosities
(eg. Figure~1 in \cite{so91}).

\begin{figure}
\centerline{
\psfig{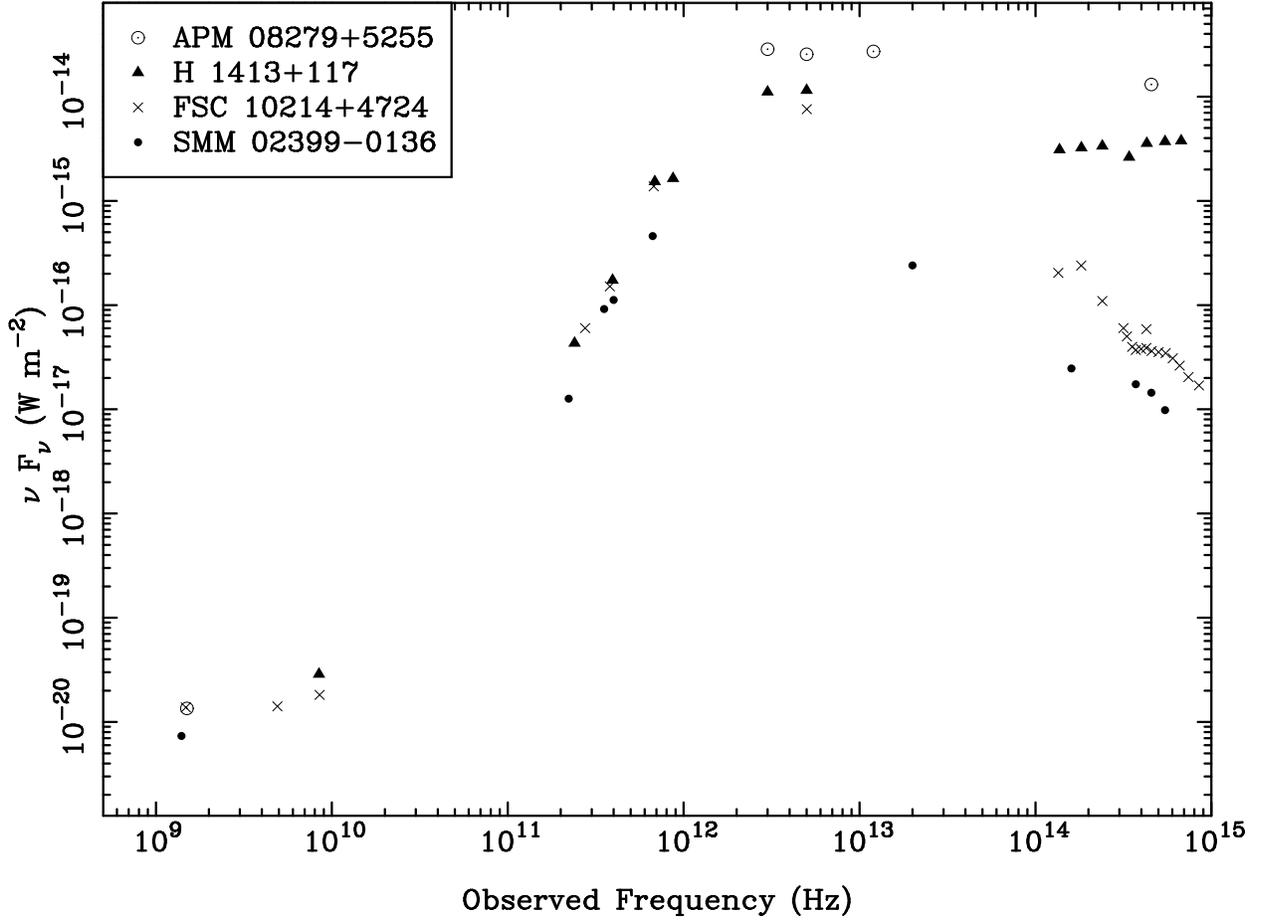}
}
\caption[]{ The spectral energy distribution of \name\ (z=3.87)
compared to the ultraluminous galaxies IRAS F10214+4724 (z=2.286), SMM
02399-0136 (z=2.8) and H1413+117 (z=2.558).  Each data--set represents
the {\it observed} flux in the form of ${\rm \nu F_{\nu}}$; only
detections have been included and upper--limits are not plotted.  Note
that H1413+117 possesses complex radio structure and the value in this
plot represents the sum of the components labelled A, B \& C$'$ by
Kayser et al. (1990).  }
\label{fig4}
\end{figure}

In Figure~\ref{fig4} we present the spectral energy distribution (SED) of
the BAL quasar, assuming that the R-band flux and the IRAS and FIRST
emission arise from the same object. The SEDs of three other hyperluminous
systems IRAS F10214+4724 at z=2.286 (\cite{ro91}), SMM 02399-0136 at z=2.8
(\cite{iv98}), and H1413+117 at z=2.558 (\cite{ba95}) are also shown.
Gravitational lensing is thought to influence the properties of all of these
systems, enhancing the observed flux by factors of $\approx$ 30, 2.5, and 30
respectively. No correction for such effects is applied to
Figure~\ref{fig4}, which presents only the observed flux values. It is
immediately apparent that, even though it is at a higher redshift, \name\
appears brighter than any of these sources in both the optical and FIR and
has a very similar SED to H1413+117. It is also apparent that, while the
flux in IRAS F10214+4724 and SMM 02399-0136 is dominated by an emission
peak in the FIR, presumably due to reprocessing by a dusty component, both
\name\ and H1413+117, while having similar FIR fluxes, also emit copiously in
the optical wavebands.  In the B-band, the SED of \name\ effectively turns
over due to absorption in the ${\rm Ly_{\alpha}}$ forest, while beyond the
Lyman limit there is little or no observed flux.

\section{Interpretation}\label{interp}

How does \name\ compare to other quasars at ${\rm z\sim4}$?
Considering a standard Friedmann--Walker universe, the distance
modulus to ${\rm z=3.87}$ is 46.02(47.12)${\rm -5\log{h}}$ for
$\Omega_o=1(0.2)$. Typically, studies of the quasar luminosity
function at high--redshift assume a $\Omega_o=1$ cosmology with ${\rm
h=0.5}$; we adopt such a cosmology to ease comparison with other
studies leading to a distance modulus of 47.52.  In a weak-lined BAL
quasar such as this, an approximation for the K-correction in the R-band 
can be obtained by assuming a power law continuum with slope 
$\alpha \approx -0.5$ leading to K$_{\rm R} = -0.86$ at a redshift
of 3.87.  Using these values implies that \name\ has an 
absolute magnitute of ${\rm M_R \approx -33.2}$. Taken at face
value, therefore, \name\ is by far the most instrinsically luminous
object known and likewise, integrating over the SED presented in 
Figure~\ref{fig4} implies that the bolometric luminosity of this system
exceeds $\sim5\times10^{15}{\rm L_{\odot}}$.

There are several possible explanations for our current observations including:

{\bf Chance alignment:} the redshift z=3.87 BAL quasar and the source of the
IRAS emission are unconnected and lie along the same line-of-sight.
Regardless of the spectral properties of the object, the likelihood ratio
for this being the correct identification given the distance from the IRAS
position and the $<$density$>$ of optical images to this magnitude is
LR$\approx$55 (eg. \cite{pp83}).  All other putative optical counterparts to
the limit of the JKT R-band data at R$=$21.4 have likelihood ratios LR$<<$1.
Furthermore, the selection of this unusual system, derived from an APM plate
color and brightness criterion, is unbiased with respect to the FIR flux, so
the chance alignment with any IRAS FSC object is highly improbable.  We
conclude that the excellent positional coincidence, extreme optical rarity
of such bright high redshift quasars and the characteristic ``dust''
spectral signature of the IRAS measurements, provide a compelling case for
the BAL quasar and the IRAS source to be either the same object, or,
somewhat more unlikely, for the
IRAS source to be related to an intervening gravitational lens that is
amplifying the flux from the QSO.

{\bf Single object:} in the simplest model the system consists of
a single source with an extremely luminous FIR and optical
flux. Absorption features in the spectrum imply the presence of other
material along the line-of-sight. Although in this model we assume
this matter does not significantly influence the observed properties
of the \name\ system which are intrinsic to the source. However, this
hypothesis is hard to reconcile with the brightness of the image
residuals of Figure~\ref{fig3}b, as the underlying extended emission
would have an absolute magnitude of ${\rm M_R > -30}$. Though not
impossible, the coincidence that such an unusually bright QSO should
reside in such an unusually bright galaxy seems somewhat contrived.

{\bf Gravitational Lensing:} Gravitational lensing by intervening material
has proven to be important in explaining the properties of a number of
high--redshift sources. The indication of the existence of significant mass
concentrations lying along the line-of-sight to this system, specifically the
\ly\ system and the two \mg\  absorbers, plus the small residuals resulting 
from the two PSF fit
to the image (Figure~\ref{fig3}c), lend support to this hypothesis. In this
scenario, the IRAS flux is most likely to be associated with the quasar, 
rather than the lensing galaxy.  

\noindent On balance, we feel that gravitational lensing is the most likely 
explanation and discuss it in more detail in the next section.

\section{The Influence of Gravitational Lensing}
\label{gravitationallensing}

Several sources which have appeared to exhibit quite phenomenal properties
have later been shown to be significantly influenced by the action of
gravitational lensing by a foreground mass distribution (e.g. the
protogalaxy candidate cB58, \cite{ye96,wi96,se98}, and the ultraluminous
IRAS source, F10214+4724, \cite{br95}).  Considering the potential
ranking of \name\ as the brightest object known, such a scenario must be
investigated.

The limited ground-based data, however, allows us to only apply the simplest
of models to the lensing potential in this system. To this end, we consider
the lens mass to be described by a singular isothermal sphere (\cite{sc92}),
a model which has been applied as a convenient description of galaxy and
cluster lenses for a number of years (e.g.~\cite{ko93}).  If the source and
the lens are then highly aligned, the source may split into two images, with
the amplification being larger the closer both images are to equal
brightness. 

In the case of perfect alignment the amplification can be (formally)
infinite.  For the isothermal sphere model, given the estimates of the
relative image brightnesses (Section~\ref{photo}), the total
amplification for a point source is estimated to be $\approx 40$.  In the
infrared the spatial extent of the emission region is likely to be larger
than in the optical and we would expect the infrared flux to be 
correspondingly less amplified.  Furthermore, due to the simplicity of the
lensing model, this should be considered only a first 
approximation to the influence of gravitational lensing.  

While the degree of magnification depends on the detailed distribution of
mass in the vicinity of the lens, the image separation is a probe of the total
mass contained within the images. Given the ground--based data, it is this
property that is considered further. We consider three individual redshifts
for the potential lensing object; firstly, $z=3.07$, coinciding with the
redshift of the \ly\ system, $z=1.81$, corresponding to the galaxy that
gives rise to the weaker of the \mg\ doublets, and $z=1.18$, the site of
the strong \mg\ absorber which is close to the statistically most
probable location of a lens in an $\Omega = 1$ universe (\cite{ko93}).
Assuming a component separation of $0\scnd4$, the resulting one--dimensional 
velocity dispersions for the isothermal model are $299\kms$, $162\kms$ and 
$130\kms$ respectively.

It is interesting to note that the small change, $\approx$40\%, in the image 
brightness, over
a four--decade period suggests that gravitational microlensing by compact
objects within the lensing body (\cite{pe97}), or along the line-of-sight
(\cite{ha96}), is unlikely to play a significant role in contributing to the
overall amplification, although this can only be strictly ruled out with
long--term monitoring of the image components.

A consequence of gravitational lensing in this system is that it 
provides two lines-of-sight through the \ly\ system. If this system is the
source of the lensing, then these paths are separated by 1.31(2.17) ${\rm
h^{-1}}$kpc in a $\Omega_o=1(0.2)$ universe. If, however, the system
corresponding to the \mg\ doublet at $z=1.18$ is the dominant lensing body,
as seems most likely, these paths are separated by 0.18(0.22) 
${\rm h^{-1}}$kpc. Spatially
resolved spectra would then probe the structure through the \ly\ system on
sub--galactic scales (e.g.~\cite{zu97}).

\section{Conclusions}\label{conclusions_section}

In this letter we have presented the discovery of a high--redshift,
ultraluminous BAL quasar, \name. Taken at face value, this system
possesses a phenomenal bolometric luminosity ${\rm\simgt 5\times
10^{15}L_{\odot}}$, making it, intrinsically, the brightest object known in
the Universe. After examining several possibilities, we suggest that the
prodigious nature of this object is most likely a consequence of
gravitational lensing by some foreground mass.  Even after correcting
the degree of amplification using a simple lensing model, the intrinsic 
luminosity of \name\ still ranks amongst the most luminous objects known.  

In addition to being optically luminous, this system appears to emit 
significant flux in the FIR, suggesting the presence of a copious quantity 
of dust.  Comparison of the FIR flux with that of F10214+4724 suggests 
an uncorrected dust mass of $\approx$6$\times10^{10}\msun$, or 
$\approx1-2\times10^9\msun$ after allowance for potential gravitational 
lensing.  Although ${\rm Ly_{\alpha}}$ emission is extremely weak, 
probably due to self-absorption by neutral gas along the line-of-sight 
within the quasar, there is no sigificant reddening of the continuum
compared to other high redshift quasars.  This suggests that the FIR
emission originates in a physically distinct region from the optical 
source, consistent with the model proposed by \cite{ba95}.  In this
context it is significant that the closest comparible object from the 
handful of other known ultraluminous systems, is the cloverleaf quasar 
H1413+117, which has a similar SED and is also a BAL quasar.

Detection of \name\ in the FIRST catalogue suggests that further followup
high resolution observations covering the entire optical to radio spectrum 
are warranted to provide vital clues to the nature of this important system.
In particular the role of gravitational lensing amplification over a wide
range of wavelengths has still to be unambiguously confirmed.

A far reaching consequence of this discovery is the possibility of similar
systems still lurking in the IRAS FSC database.  So far three ultraluminous
high redshift galaxies with z$>$2 have been detected in the IRAS FSC. 
Systematic spectroscopic campaigns have only probed a mere 10\% of the 
galaxies in the FSC and uncovered F10214+4724.  Both H1413+117 and
\name\ were discovered during other observing campaigns and subsequently
found to have IRAS FSC fluxes.  All three objects appear to be present in 
the IRAS FSC because of significant amplification by foreground lensing
systems, raising the question of how many more lensed high redshift
galaxies remain to be discovered in the IRAS database.

\section{Acknowledgements}\label{ackers}
We would like to thank David Schlegel for kindly taking the direct images 
on the JKT for us. The anonymous referee is thanked for useful comments.

\newpage

\newpage

\newpage

\newpage

\clearpage
 
\begin{deluxetable}{cccc}
\footnotesize
\tablecaption{ 
IRAS and FIRST flux measures of APM 08279+5255.
\label{table}}
\tablewidth{0pt}
\tablehead{
\colhead{Catalog} & \colhead{$\lambda$} &  \colhead{Flux} & 
\colhead{Error}
} 
\startdata
IRAS   & 12$\rm \mu m$ & \hspace{-0.15in} $<$0.1010  Jy & $-$  \nl
IRAS   & 25$\rm \mu m$ &       0.2261  Jy & $16\%$\nl
IRAS   & 60$\rm \mu m$ &       0.5111  Jy & $10\%$\nl
IRAS   &100$\rm \mu m$ &       0.9511  Jy & $24\%$\nl
FIRST  & 21$\rm    cm$ &       0.92   mJy & $10\%$\nl
\enddata
\end{deluxetable}


\newpage

\begin{thebibliography}{}

\bibitem[Barvainis et al.\ 1992]{ba92} Barvainis, R., Antonucci, R. \&
Coleman, P., 1992, \apj\, 399, L19

\bibitem[Barvainis et al.\ 1994]{ba94}
  Barvainis, R., Tacconi, L., Antonucci, R. \& Alloin, D., 
  1994, \nat\, 371, 586

\bibitem[Barvainis et al.\ 1995]{ba95}
  Barvainis, R., Antonucci, R., Hurt, T., Coleman, P. \& Reuter, H.-P.,
  1995, \apj\, 451, 9 

\bibitem[Broadhurst \& Lehar 1995]{br95}
  Broadhurst, T. \& Lehar, J., \apj\, 450, 41

\bibitem[Hawkins 1996]{ha96}
  Hawkins, M. R. S.,
  1996, \mnras\, 278, 787

\bibitem[Hook et al.\ 1994]{ho94}
  Hook, I. M., McMahon, R. G., Boyle, B. J. \& Irwin, M. J.,
  1994, \mnras\, 269, 305

\bibitem[Guilloteau et al.\ 1997]{gu97}
  Guilloteau, S., Omont, A., McMahon, R. G., Cox, P. \& Petitjean, P.,
  1997, \aap, 328, 1
  
\bibitem[Ivison et al.\ 1998]{iv98}
  Ivison, R. J., Smail, I., Le Borgne, J.-F., Blain, A. W., Kneib, J.-P.,
  B\'{e}zzcourt, J., Kerr, T. H. \& Davies, J. K.,
  1998, preprint, {\it astro-ph/9712161}

\bibitem[Kayser et al.\ 1990]{ka90}
  Kayser, R., Surdej, J., Condon, J. J., Kellermann, K. I., Magain, 
  P., Remy, M. \& Smette, A.,
  1990, \aap, 364, 15

\bibitem[Kochanek 1993]{ko93}
  Kochanek, C. S., 1993, \apj\, 419, 12

\bibitem[Lasker et al.\ 1988]{lask86}
  Lasker, B. M., et al. 1988, \apjsupp\, 68, 1

\bibitem[Magain et al.\ 1988]{ma88}
  Magain, P., Surdej, J., Swings, J.-P., Borgeest, U. \& Kayser, R.,
  1988, \nat\, 334, 325

\bibitem[Oliver et al.\ 1996]{ol96}
  Oliver, S. J., et al. 1996, \mnras\, 280, 673

\bibitem[Omont et al.\  1996]{om96}
  Omont, A., Petitjean, P., Guilloteau, S., McMahon, R. G.,
  Solomon, P. M. \& Pecontal, E.,
  1996, \nat\, 382, 428

\bibitem[Perna \& Loeb 1997]{pe97}
  Perna, R. \& Loeb, A., 
  1997, \apj\, 489, 489
  
\bibitem[Prestage \& Peacock 1983]{pp83}
  Prestage, R., \& Peacock, J., 1983, \mnras\, 204, 355

\bibitem[Rowan-Robinson et al.\ 1990]{ro90}
  Rowan-Robinson, M., Lawrence, A., Saunders, W., Crawford, J., Ellis, R., 
  Frenk, C. S., Parry, I., Xiaoyang, X., Allington-Smith, J., Efstathiou, G. 
  \& Kaiser, N.
  1990, \mnras\, 247, 1

\bibitem[Rowan-Robinson et al.\ 1991]{ro91}
  Rowan-Robinson, M., Broadhurst, T., Oliver, S.J., Taylor, A.N.,
  Lawrence, A., McMahon, R.G., Lonsdale, C.J., Hacking, P.B. \&
  Conrow, T., 1991, \nat\, 351, 719
  
\bibitem[Schneider et al.\ 1992]{sc92}
  Schneider, P., Ehlers, J. \& Falco, E. E., 
  1992, {\it Gravitational Lenses}, Springer-Verlag Press, Berlin
  
\bibitem[Seitz et al.\ 1988]{se98}
  Seitz, S., Saglia, R. P., Bender, R., Hopp, U., Belloni, P. \& Ziegler, B.,
  preprint, astro-ph/9706023

\bibitem[Sopp \& Alexander 1991]{so91}
  Sopp, H. M. \& Alexander, P.,
  1991, \mnras\, 251, 14

\bibitem[Stetson 1987]{st87}
  Stetson, P.,
  1987, \pasp\, 99, 191

\bibitem[Storrie-Lombardi et al.\ 1996]{sl96}
  Storrie-Lombardi, L.J., McMahon, R.G., Irwin, M.J., \& Hazard, C.,
  1996, \apj\, 468, 121

\bibitem[Strauss et al.\ 1990]{st90}
  Strauss, M. A., Davis, M., Yahil, A. \& Huchra, J. P.,
  1990, \apj\, 361, 49 

\bibitem[Totten \& Irwin 1998]{ed98}
  Totten, E. J., \& Irwin, M. J., 
  1998, \mnras\, 294, 1

\bibitem[Turnshek 1988]{tu88}
  Turnshek, D. A., 
  1988, in {\it QSO Absorption Lines: Probing the Universe}, 
  Eds. Blades, J. C., Turnshek, D. A. \& Norman, C. A., 
  Cambridge University Press.
 
\bibitem[Weymann 1995]{we95}
  Weymann, R.,
  1995, in {\it QSO Absorption Lines}, 
  Ed. Meylan, G., Springer Verlag.

\bibitem[Williams \& Lewis 1996]{wi96}
  Williams, L. L. R. \& Lewis, G. F., 1996, \mnras\, 281, L35
 
\bibitem[Yee et al.\ 1996]{ye96}
  Yee, H. K. C., Ellingson, E., Bechtold, J., Carlberg, R. G. 
  \& Cuillandre, J.-C., 1996, \aj\, 111, 1883 

\bibitem[Zuo et al.\ 1997]{zu97}
  Zuo, L., Beaver, E. A., Burbidge, E. M., Cohen, R. D.,
  Junkkarinen, V. T. \& Lyons, R. W.,
  1997, \apj\, 477, 568
 
\end{thebibliography}
\end{document}